\documentclass{elsart}
\usepackage{graphicx}
\usepackage{amsmath,amssymb}

\begin{document}
\begin{frontmatter}

\title{Size effects and conductivity of ultrathin Cu films}
\author[1,2]{Dmitry V.~Fedorov},
\author[2]{Peter Zahn\corauthref{cor1}}, and
\ead{zahn@physik.uni-halle.de}
\corauth[cor1]{Corresponding author}
\author[2]{Ingrid Mertig}
\address[1]{Physical-Technical Institute, 
Ural Branch of Russian Academy of Sciences, 426001 Izhevsk, Russia}
\address[2]{Martin-Luther-Universit\"at Halle-Wittenberg, Fachbereich Physik,
Fachgruppe Theoretische Physik, D-06099 Halle, Germany}

\begin{abstract}
We propose a model for the description of the transport properties of metallic films
on a large scale of slab thickness. This model is based on solving the linearized
Boltzmann equation in relaxation-time approximation using {\it ab initio} calculations 
within the framework of the density functional theory. The expression for the relaxation time
is derived from the microscopic treatment of the scattering processes and provides the
correct thickness dependence for very thin as well as very thick films. The method is applied 
to the calculation of the in-plane conductivity and the Drude-type plasma frequency
of thin Cu(001) films in the thickness range between 1 and 32 monolayers.
\end{abstract}
\end{frontmatter}

\section{Introduction}
The electronic structure of ultrathin metallic films is of great scientific
interest. Their transport properties can substantially deviate from the bulk
behaviour owing to classical and quantum size effects and the existence
of new types of electronic states (e.g. surface states). Therefore, the study of these
systems is very important for the microscopic understanding of transport in
nanoscaled electronic devices. 

Classical size effects in the conductivity as explained by the Fuchs-Sondheimer theory 
occur if the slab thickness $d$ is comparable to the bulk mean free path \cite{Fuchs38,Sond52}.
In the limit of small $d$ compared to the mean free path this model predicts a linear 
thickness dependence of the slab conductivity. The application of the model to this case 
is quite questionable, since a three-dimensional electron motion is assumed. Nevertheless,
this behaviour is confirmed by ultrathin film experiments.

When the thickness $d$ approaches the Fermi wavelength $\lambda_F$ quantum size effects (QSE) 
occur which cause oscillations in typical features of the electronic structure, e.g.
density of states at the Fermi level, as a function of slab thickness. Regarding 
the transport properties Trivedi and Ashcroft predicted a saw-tooth like oscillation 
of the in-plane conductivity with thickness \cite{Trivedi88}. In their free-electron 
model the oscillation period is equal to half $\lambda_F$. In reality, the
situation is more complicated because of the discrete increase of the film thickness
as well as the change in the electronic structure during growth. There are several
interesting experiments which demonstrate QSE. For example, 
the observation of QSE in the dependence of the Hall coefficient and the specific
conductivity on the thickness of ultrathin Pb(111) films \cite{Jaloch96}, the step
height oscillations during layer-by-layer growth of Pb on Ge(001) \cite{Crottini97},
and the manifestation of quantum-well states in angle-resolved photoemission from Ag 
films \cite{Paggel99} are worthy of mention.

Recently, the conductivity of ultrathin 
Pb(111) films consisting of $N=1\ldots$8 monolayers (ML) was studied by means of
{\it ab initio} calculations using the full-potential linearized
augmented plane-wave method \cite{Vilfan03}. The results obtained are in a good agreement with the
experiment \cite{Vilfan02}, and showed a correlation between QSE oscillation in the
conductivity and in the density of states at the Fermi level. 

In an earlier paper, we have investigated the evolution of the Co/Cu multilayer conductivity 
during growth using the Boltzmann equation in relaxation-time approximation \cite{Zahn02}.
The transport properties of Cu(001) films of 2 to 5 free-standing ML have
already been studied. 
Now this technique is extended to Cu slabs in the thickness range between
1 and 32 ML. 
We propose a model for the relaxation time of electrons
caused by elastic impurity scattering in slabs which
allows us to describe the conductivity in the limit of both very thin and
very thick films.  

\section{The model}
The electronic structure of the considered systems is calculated
self-consistently using a screened Korringa-Kohn-Rostoker (SKKR) method
\cite{Zahn98,And92,Szun94,Zeller95,Nikos02}.
For our calculations a screening potential with a barrier height of 4 Ry 
is used and the screened structure constants include coupling to the next four 
nearest neighbours. Spherical potentials in the atomic sphere
approximation (ASA) are used. The calculations employ the exchange-correlation 
potential in the LDA of Vosko, Wilk, and Nusair \cite{Vosko80}. An angular 
momentum cut-off at $l_{max} = 3$ is used for the Green's function, thus 
implying a cut-off for the charge-density components at $2 l_{max}= 6$. 
Due to the slab geometry the states at the Fermi level are described by lines
in the two-dimensional (2D) Brillouin zone (BZ). The Fermi surface integrals have
to be replaced by line integrals \cite{Zahn02}. In the considered geometry
the irreducible part equals one eighth of the 2D BZ and contains in our
calculations about 28000 ${\bf k}_{||}$-points to reach convergence. 
For comparison, the Fermi surface integration for the bulk Cu
system is performed on about 350000 ${\bf k}$-points in the irreducible part
of the fcc BZ. We neglect the lattice relaxations at the surfaces. 
The fcc lattice constant is taken equal to A=6.76 a.u. \cite{Moruzzi78}.

The transport is treated quasiclassically by solving the linearized Boltzmann equation which 
in the anisotropic relaxation-time approximation gives the following expression for the 
components of the conductivity tensor $\hat \sigma$ of a crystal \cite{Zahn95,Mertig99}
$$
\sigma_{ij}= \frac{2 e^2}V \sum\limits_ k 
\delta (E_k-E_F) v_k^i v_k^j\tau_k \ ,
\eqno (1)
$$
where $e$ is the electron charge, $V$ is the system volume, $E_F$ the Fermi energy, $v_k^i$ 
the component of the group velocity with a Cartesian coordinate $i$ of an electron in the 
band $n$ with crystal momentum vector ${\bf k}$ and  relaxation time $\tau_k$ at the 
energy $E_k$ ($k$ is a shorthand notation for ${\bf k}$ and $n$). The factor 2 is due 
to the spin degeneracy. For a cubic crystal and a 2D lattice with square symmetry
the transport properties are isotropic $\sigma _{ij} =\sigma \delta_{ij}$.

In the case of film geometry the relaxation time $\tau_{film}$ has to be
determined by the scattering at 
impurities inside of the slab as well as on surface imperfections
$$
\frac 1{\tau_{film}} = \frac 1{\tau^b} + \frac 1{\tau^s}\ .
\eqno (2)
$$ 
Since our calculations 
are performed at $T=0 K$, the inelastic scattering by phonons is neglected.
First, we consider the scattering rate $(\tau^b)^{-1}$ caused by
bulk-like defects inside the slab. 
The microscopic transition probability 
for an electron to be scattered from a state $k$ into a state $k^{\prime}$, is given by Fermi's 
Golden Rule \cite{Zahn98,Mertig99} 
$$
P_{kk^{\prime}}=\frac{2\pi}{\hbar} |T_{kk^{\prime}}|^2\delta(E_k-E_{k^{\prime}}) \ ,
\eqno (3)
$$
where the transition matrix $T_{kk^{\prime}}$ is defined by
$$
T_{kk^{\prime}}=<{\mathaccent"17 {\Psi}_k}|\Delta V|\Psi_{k^{\prime}}> \ .
\eqno (4)
$$
Here $\Delta V$ denotes the perturbation of the potential, ${\mathaccent"17 {\Psi}_k}$ and 
$\Psi_{k^{\prime}}$ are the unperturbed and perturbed Bloch states of the system, respectively. 
Within the assumption of $\delta$-scatterers \cite{Zahn98} with the scattering strength 
$t_b$ at all impurity sites ${\bf r}_i$ in the slab 
$$
\Delta V({\bf r})=\sum\limits_{i=1}^{M_b} t_b\delta({\bf r}-{\bf r}_i)\ ,
\eqno (5)
$$
the relaxation time $\tau_k^b$ is given by the expression \cite{Zahn98} 
$$
\frac 1{\tau_k^b} = \sum\limits_{k^{\prime}} P_{kk^{\prime}} =
\frac{2\pi}{\hbar} t_b^2\sum\limits_{i=1}^{M_b}
|{\mathaccent"17 {\Psi}_k} ({\bf r}_i)|^2\  
n({\bf r}_i, E_F)\ \ . 
\eqno (6)
$$
The number $M_b$ denotes the number of bulk-like defects in the slab. 
Equation (6) includes the probability amplitude of a state $k$ at the impurity position 
$|{\mathaccent"17 {\Psi}_k} ({\bf r}_i)|^2$ and the local density of
states at Fermi level $n({\bf r}_i, E_F)$.
Further simplifications can be achieved by averaging the scattering rate (6) over all
states $k$ at the Fermi energy $E_F$ and over the configurations of the impurity sites
${\bf r}_i$. We assume the same probability amplitude of perturbed 
Bloch state at all impurity sites in the film. Introducing the bulk density
$\varrho_b=M_b /S d$ of scattering centres, with $S$ the area of the film and $d$ 
the slab thickness, the averaged relaxation time is given by
$$
\frac 1{\tau^b}= \frac{2\pi n_{film}(E_F) t_b^2\varrho_b} {\hbar S_0 d}\ ,
\eqno (7)
$$
where $n_{film}(E_F)$ is the total film density of states at the Fermi level (per slab unit 
cell). The area of the film unit cell is given by the film area
divided by the number of unit cells in the slab $S_0=S / N_{cell}$. 

Now, to describe the second term in Eq.(2) we introduce the additional number $M_s$ of the
surface imperfections on the slab surfaces with the surface density $\varrho_s =M_s / S$.
Similar to the above derivation, an expression for $\tau^s$ is obtained
$$
\frac 1{\tau^s}= \frac{2\pi n_{surf}(E_F) t_s^2\varrho_s} {\hbar S_0 D_s d}\ .
\eqno (8)
$$
Here $n_{surf}(E_F)$ is the local density of states in the surface region at the Fermi 
level (per slab unit cell), $t_s$ is the parameter for the scattering strength of the surface
imperfections. $D_s$ is the thickness of the surface region which contains the surface
defects.  

Introducing the defects of scattering strength $t_b$ in a bulk material, the bulk
scattering rate is obtained similar  
$$
\frac 1{\tau_{bulk}}=\frac{2\pi n_{bulk}(E_F) t_b^2\varrho_b}{\hbar S_0 D}\ ,
\eqno (9)
$$
where $n_{bulk}(E_F)$ is the density of states at the Fermi level per bulk unit cell. 
The volume of the bulk unit cell is expressed by $S_0 D$ with
the interlayer distance $D=d/N$, where $N$ denotes the number of 
monolayers in the slab.

Now, introducing the dimensionless function 
$$
\gamma(N) = \frac{\tau_b N}{\tau_s} =
\frac{t_s^2\varrho_s n_{surf}(E_F) N}{t_b^2\varrho_b n_{film}(E_F) D_s}\ ,
\eqno (10)
$$ 
and using equations (7), (8), and (9), the relaxation time in the slab is obtained by
$$
\tau_{film}=\frac{N n_{bulk}(E_F)}{n_{film}(E_F)}\cdot\frac{\tau_{bulk}}{(1+\gamma(N)/N)}\ .
\eqno (11)
$$

The function $\gamma(N)$ has a clear physical meaning. It defines the ratio 
between the strength of the electron scattering at surface imperfections and at impurities
inside of the slab. By virtue of the proportionality of  
$n_{film}(E_F)$ to $N$ the function $\gamma(N)$ is approximated in
this work by a constant value $\gamma$, 
even for the case of small thicknesses. Moderate changes in $\gamma(N)$ for small thicknesses 
may change the results quantitatively, but the general trend remains unchanged. 
In the limit of large $N\gg \gamma $ the relaxation time in the film
$\tau_{film}$ reaches the bulk value $\tau_{bulk}$. In the limit of
very thin films the expression (11) is equivalent to the one used in the ref.\cite{Vilfan03}
and that derived from a free-electron model assuming elastic and diffuse surface scattering 
\cite{Calecki90,Camblong99}. 
The authors of \cite{Vilfan03} considered ultrathin Pb(111) films ($N = 1\ldots$8 ML) 
with only one adjustable parameter like $t_s^2\varrho_s$ in Eq.(8), and they obtained 
a linear dependence of the conductivity on the slab thickness which is in a good agreement 
with the experiment \cite{Vilfan02} for the considered thickness range. In addition, we have introduced 
$t_b^2\varrho_b$ to treat thicker slabs. Their conductivity is appreciably 
determined by the properties of the bulk region of the slab 
and does not exhibit the linear thickness dependence (see, e.g. \cite{Jaloch96}). 
Normalizing the film conductivity $\sigma_{film}$ to the bulk conductivity 
$\sigma_{bulk}$ it is possible to keep one adjustable parameter $\gamma(N)$ only. 
 
Finally, performing the integration in Eq.(1) over the Fermi surface (for the slab --- the Fermi 
lines) \cite{Ziman72}, we obtain in case of cubic symmetry
$$
\frac{\sigma_{film}}{\sigma_{bulk}} = \frac {3\pi n_{bulk}(E_F)\sum_n\int dl_n |v_n (E_F)|}
{n_{film}(E_F) D (1+\gamma(N)/N) \sum_n\int dS_n |v_n (E_F)|} \ ,
\eqno (12)
$$
where $dl_n$ and $dS_n$ are differential Fermi line and Fermi surface elements,
respectively.

In case of constant relaxation times $\tau_{film}\equiv\tau_{bulk}$ for all slab thicknesses 
(like in ref.\cite{Zahn02}) the ratio of the conductivities
$\sigma_{film} / \sigma_{bulk}$ can be transformed to the ratio
$\omega_{{\text P}, film}^2 / \omega_{{\text P}, bulk}^2$ 
$$
\frac{\omega_{{\text P}, film}}{\omega_{{\text P}, bulk}} =
\left( \frac {3\pi\sum_n\int dl_n |v_n (E_F)|}
{N D\sum_n\int dS_n |v_n (E_F)|}\right)^{1/2} \ ,
\eqno (13)
$$
where $\omega_{{\text P}, film}$ and $\omega_{{\text P}, bulk}$ are the slab and the bulk 
Drude-type plasma frequencies \cite{Fahsold03}, respectively. This ratio can be extracted 
from infrared transmission experiments \cite{Fahsold}.

\section{Results}
In Fig.1 we show the total slab density of states at the Fermi level $n_{film} (E_F)$ determined from 
the one-particle energy spectrum $E_n ({\bf k})$. In addition to the peak for films of 2 and 3
ML in thickness large densities at the Fermi level are obtained for the thicknesses of 9, 17, and 30 ML.
In the asymptotic case of very large $N$ the oscillations in $n (E_F)$ are governed by Fermi
surface nesting vectors along the (001) direction, which are known to dominate the 
oscillations of interlayer exchange coupling \cite{Bruno91}. In the investigated thickness 
range the oscillations with a period of 2.6 ML could be identified by Fourier
analysis.
This oscillation corresponds to the short-period oscillation of interlayer
exchange coupling \cite{Bruno91}.

The evolution of electrical conductivity with film thickness, Eq.(12), 
is presented in Fig.2 with different values 
of the parameter $\gamma $ for the strength of the surface scattering.  
In principle, it is possible to get the value of $\gamma$ (or the functional dependence 
$\gamma (N)$) from a comparison with experimental data. Unfortunately,   
experimental data concerning the conductivity during growth of very thin Cu
slabs are missing.  
The reason is the prefered initial growth mode of Cu on insulating substrates, which
forms islands instead of a continous film \cite{Cu_growth}. 
Reasonable values of $\gamma$ 
can roughly be estimated from a comparison with the experimental data for the conductivity 
of Fe \cite{Fahsold03} and Pb \cite{Jaloch96} films. The ratio $\sigma_{film} / \sigma_{bulk}$ 
is about 0.6 for an Fe(001) slab with 31 ML ($\approx$ 4.5 nm) thickness, and 
$\sigma_{film} / \sigma_{bulk}$ is about 0.3 for a Pb(111) film with 18 ML ($\approx$ 5.1 nm) 
thickness, assuming a bulk Pb conductivity of $0.213\ (\mu\Omega cm)^{-1}$ \cite{Ashcroft}. 
Of course, these data show the strong dependence on the material. Nevertheless, the range 
of $\gamma$ should be between 10 and 100 to describe the existing experimental 
situations. In this range, strong QSE's can be recognized from Fig.2, which
are clearly correlated to the behaviour of $n (E_F)$ in Fig.1. 

Figure 3 shows the results for the ratio of plasma frequencies
$\omega_{{\text P}, film} / \omega_{{\text P}, bulk}$. 
From this picture a $1/N$ behaviour for large $N$ is evident.
In the framework of a model of free electrons in a box with the height 
$d=N D$ and Fermi wave vector $k_F$ the asymptotic behaviour for large $N$ is obtained as
$(\omega_{{\text P}, film}/\omega_{{\text P}, bulk})^2\approx 1- 3\pi/4 k_F d-\pi^2/4 k_F^2 d^2$. 
The same expression was found for the asymptotic behaviour of the charge density in 
ref.\cite{Trivedi88}, since $\omega_{\text P}^2$ is proportional 
to the charge density in the free-electron model. 

The strong increase of $\omega_{{\text P}, film}$ for small $N$ is related to the 
increase of the Fermi velocity caused by the large contribution of fast
surface states. This is demonstarted in Fig.4 presenting
the Fermi velocity $v_F=\langle|v_k(E_F)|\rangle_k$ averaged over all states $k$. 
The ratio of $v_F$ for a 32 ML Cu(001) slab to the averaged Fermi 
velocity of Cu bulk is about 0.766. It is obvious, that this
ratio does not reach one even for large $N$, 
since the Fermi velocity in the slab geometry corresponds to the projection of the bulk
Fermi velocity into a plane perpendicular to the slab growth direction. The free-electron 
model gives a ratio of $\pi /4\approx 0.785$. The strong similarities
of our results calculated for Cu slabs with those found 
from a free-electron model are direct consequences of the free-electron character 
of the Cu Fermi surface \cite{Mertig99}.

The qualitative behaviour of the $\omega_{{\text P}, film} / \omega_{{\text P}, bulk}$ 
is very close to data obtained from infrared transmission spectroscopy performed during 
the evaporation of Fe on MgO(001) in ultrahigh vacuum 
(see fig.5 in ref.\cite{Fahsold}). Although we have to note, that in the 
experiment the strong increase of $\omega_{{\text P}, film}$ for very 
thin slabs can be due to a strong influence from the morphology and the large
roughness of the films just
above the percolation threshold. We should mention also, that 
$\omega_{{\text P}, film}$ does not saturate to a constant value after 4-5 ML as we reported 
in an earlier paper \cite{Zahn02}. This conclusion was drawn based on the
results of Cu slabs with thickness of from 2 to 5 free-standing monolayers.

\section{Conclusion}
We present a model for the description of the transport properties of thin metallic films  
based on the solution of the linearized Boltzmann equation with a quite simple and at the same time 
reasonable assumption for the relaxation time. This approximation for the relaxation time 
derived from the microscopic treatment of the scattering processes has only one adjustable 
parameter, and is valid in the limits of very thin as well as very thick films. 
The qualitative behaviour of the slab conductivity found in this approach is close to the
experimental data for thin metallic films. For very thin films we see a linear dependence
of the conductivity on the thickness which was expected from the Fuchs-Sondheimer theory.
In addition, the influence of QSE's is superimposed and causes the dips on the
otherwise monotonic curve. 
The calculated Drude-type plasma frequency shows
the dominance of the surface electronic structure for thicknesses smaller than 5 ML . 
So, both classical and quantum size effects, as well as surface states manifest in our results.

The method applied in this paper can be used for metallic films with arbitrary thickness
limited by the computer power only.  

\section*{ACKNOWLEDGMENTS}
This work has been supported by the Deutsche Forschungsgemeinschaft, and by INTAS
(Grant No. 2001-0386). Valuable and encouraging discussions with Gerhard Fahsold 
and Annemarie Pucci are appreciated.

\clearpage

\begin{figure}[!htb]
\includegraphics[width=.9\textwidth]{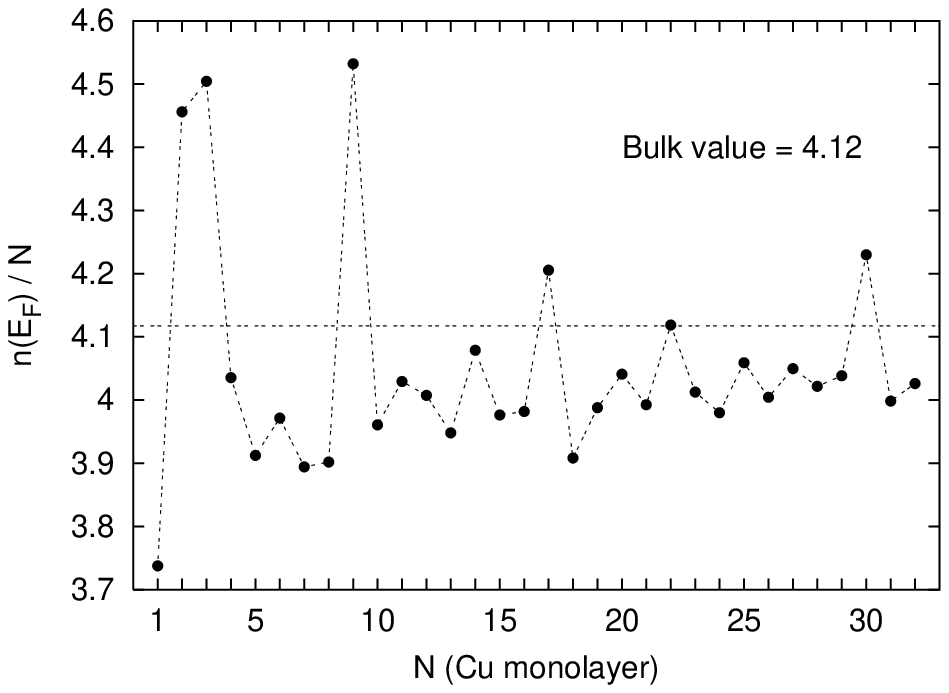}
\caption{Density of states at the Fermi level per atom 
as a function of Cu layer thickness.}
\end{figure}

\begin{figure}[!htb]
\includegraphics[width=.9\textwidth]{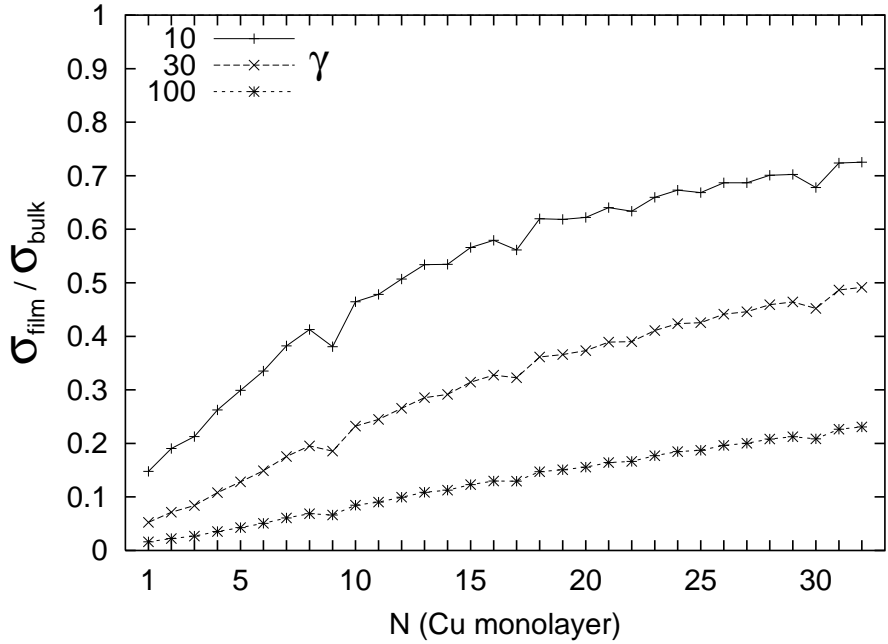}
\caption{Film conductivity (in units of the bulk conductivity) 
as a function of Cu layer thickness.}
\end{figure}

\begin{figure}[!htb]
\includegraphics[width=.9\textwidth]{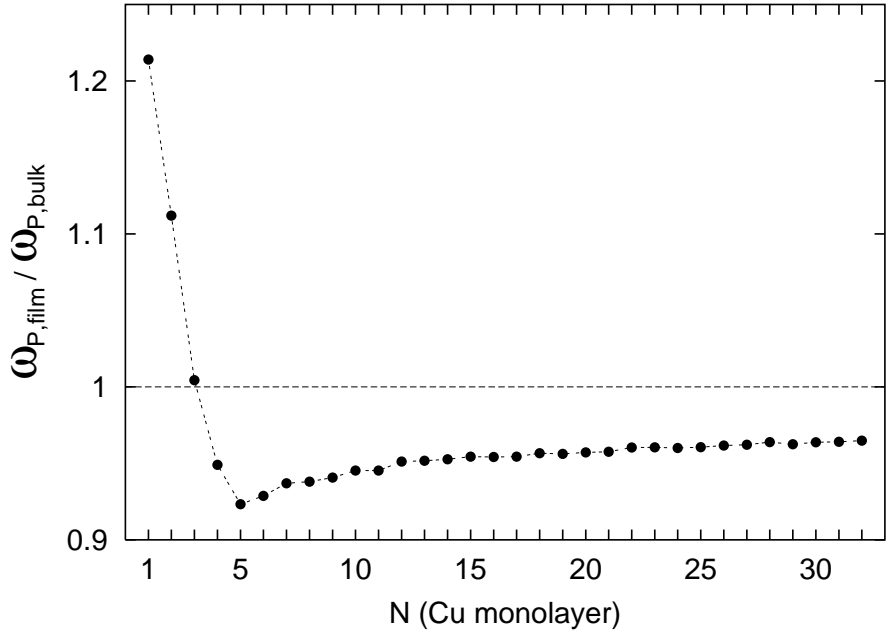}
\caption{Film plasma frequency (in units of the bulk plasma frequency) 
as a function of Cu layer thickness.}
\end{figure}

\begin{figure}[!htb,floatfix]
\includegraphics[width=.9\textwidth]{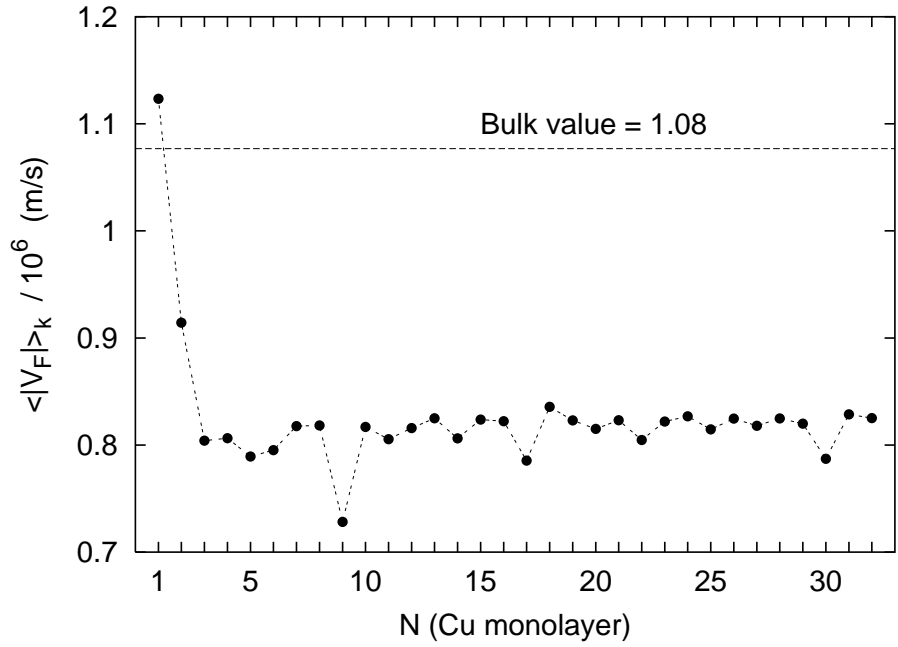}
\caption{Averaged Fermi velocity of the Cu slabs as a function of thickness.
The dashed line marks the bulk value.}
\end{figure}

\end{document}